# Logic Design

Issam W. Damaj, Dhofar University

## Introduction

Over the years, digital electronic systems have progressed from vacuum-tube to complex integrated circuits, some of which contain millions of transistors. Electronic circuits can be separated into two groups, digital and analog circuits. Analog circuits operate on analog quantities that are continuous in value and in time, while digital circuits operate on digital quantities that are discrete in value and time (1).

Analog signals are continuous in time besides being continuous in value. Most measurable quantities in nature are in analog form, for example, temperature. Measuring around the hour temperature changes is continuous in value and time, where the temperature can take any value at any instance of time with no limit on precision but on the capability of the measurement tool. Fixing the measurement of temperature to one reading per an interval of time and rounding the value recorded to the nearest integer will graph discrete values at discrete intervals of time that easily could be coded into digital quantities. From the given example, it is clear that an analog-by-nature quantity could be converted to digital by taking discrete-valued samples at discrete intervals of time and then coding each sample. The process of conversion is usually known as analog-to-digital conversion (A/D). The opposite scenario of conversion is also valid and known as digital-to-analog conversion (D/A). The representation of information in a digital form has many advantages over analog representation in electronic systems. Digital data that is discrete in value, discrete in time, and limited in precision could be efficiently stored, processed and transmitted. Digital systems are said practically to be more noise immune as compared to analog electronic systems due to the physical nature of analog signals. Accordingly, digital systems are more reliable than their analog counterpart. Examples of analog and digital systems are shown in Figure 1.

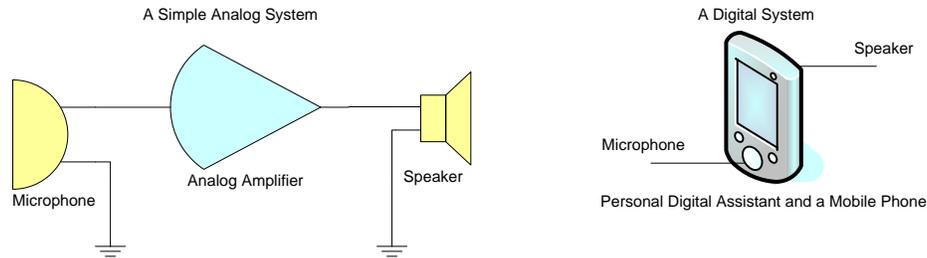

**Figure 1. A simple analog system and a digital system; the analog signal amplifies the input signal using analog electronic components. The digital system can still include analog components like a speaker and a microphone, the internal processing is digital.**

## A Bridge between Logic and Circuits

Digital electronic systems represent information in digits. The digits used in digital systems are the *0* and *1* that belong to the *binary* mathematical number system. In logic, the *1* and *0* values correspond to *True* and *False*. In circuits, the *True* and *False* could be thought of as *High* voltage and *Low* voltage. These correspondences set the relationships among logic (*True* and *False*), *binary* mathematics (*0* and *1*), and circuits (*High* and *Low*).

Logic, in its basic shape, deals with reasoning that checks the validity of a certain proposition - a proposition could be either *True* or *False*. The relationship among logic, *binary* mathematics, and circuits enables a smooth transition of processes expressed in propositional logic to *binary* mathematical functions and equations (*Boolean* algebra), and to digital circuits. A great scientific wealth exists that strongly supports the relationships among the three different branches of science that lead to the foundation of modern digital hardware and logic design.

*Boolean* algebra uses three basic logic operations *AND*, *OR*, and *NOT*. The *NOT* operation if joined with a proposition *P* works by negating it; for instance, if *P* is *True* then *NOT P* is *False* and vice versa. The operations *AND* and *OR* should be used with two propositions, for example, *P* and *Q*. The logic operation *AND*, if applied on *P* and *Q* would mean that *P AND Q* is *True* only when both *P* and *Q* are *True*. Similarly, the logic operation *OR*, if applied on *P* and *Q*, would mean that *P OR Q* is *False* only when *P* and *Q* are *False*. Truth tables of the logic operators *AND*, *OR*, and *NOT* are shown in Figure 2.a. Figure 2.b shows an alternative representation of the truth tables of *AND*, *OR*, and *NOT* in terms of *0s* and *1s*.

## Combinational Logic Circuits

Digital circuits implement the logic operations *AND*, *OR*, and *NOT* as hardware elements called "gates" that perform logic operations on binary inputs. The *AND*-gate performs an

AND operation, an *OR*-gate performs an *OR* operation, and an *Inverter* performs the negation operation *NOT*. Figure 2.c shows the standard logic symbols for the three basic operations. With analogy from electric circuits, the functionality of the *AND* and *OR* gates are captured as shown in Figure 3. The actual internal circuitry of gates is built using transistors; two different circuit implementations of inverters are shown in Figure 4. Examples of *AND*, *OR*, *NOT* gates integrated circuits (*ICs*) are shown in Figure 5. Besides the three essential logic operations, there are four other important operations - the *NOR* (*NOT-OR*), *NAND* (*NOT-AND*), Exclusive-*OR* (*XOR*) and Exclusive-*NOR* (*XNOR*).

| Input P | Input Q | Output: P AND Q | Input P | Input Q | Output: P OR Q | Input X | Output: NOT P |
|---|---|---|---|---|---|---|---|
| False | False | False | False | False | False | False | True |
| False | True | False | False | True | True | True | False |
| True | False | False | True | False | True | | |
| True | True | True | True | True | True | | |

(a)

| Input P | Input Q | Output: P AND Q | Input P | Input Q | Output: P OR Q | Input X | Output: NOT P |
|---|---|---|---|---|---|---|---|
| 0 | 0 | 0 | 0 | 0 | 0 | 0 | 1 |
| 0 | 1 | 0 | 0 | 1 | 1 | 1 | 0 |
| 1 | 0 | 0 | 1 | 0 | 1 | | |
| 1 | 1 | 1 | 1 | 1 | 1 | | |

(b)

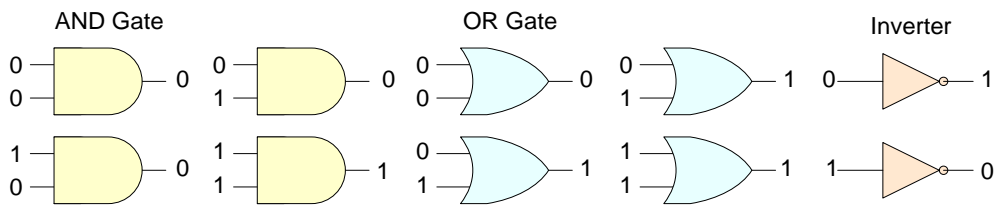

(c)

**Figure 2. (a) Truth tables for AND, OR, and Inverter. (b) Truth tables for AND, OR, and Inverter in binary numbers, (c) Symbols for AND, OR, and Inverter with their operation.**

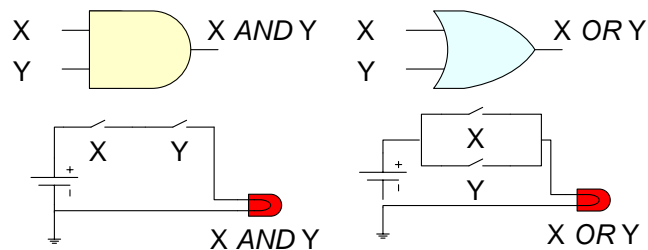

**Figure 3. A suggested analogy between *AND* and *OR* gates and electric circuits.**

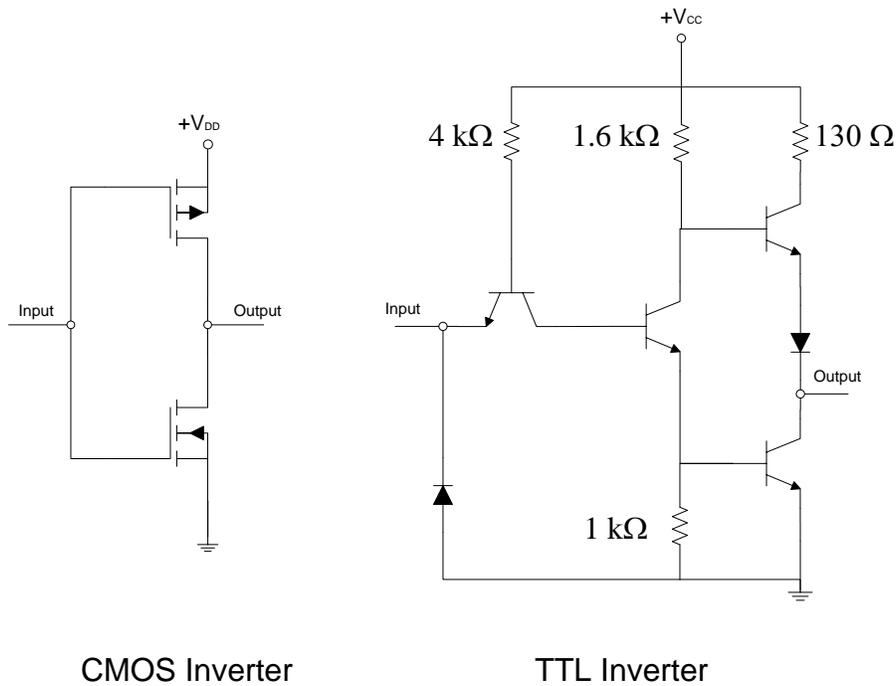

**Figure 4. Complementary Metal-oxide Semiconductor (*CMOS*) and Transistor-Transistor Logic (*TTL*) Inverters.**

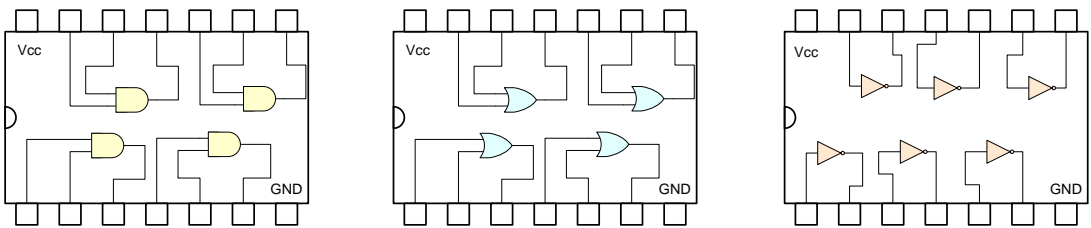

**Figure 5. The 74LS21 (*AND*), 74LS32 (*OR*), 74LS04 (*Inverter*) *TTL* ICs.**

A combinational logic circuit is usually created by combining gates together to implement a certain logic function. A combinational circuit produces its result upon application of its input(s). A logic function could be a combination of logic variables, such as *A*, *B*, *C*, etc. Logic variables can take only the values *0* or *1*. The created circuit could be implemented using *AND-OR-Inverter* gate-structure or using other types of gates. Figure 6.a shows an example combinational implementation of the following logic function *F(A, B, C)*:

*F(A, B, C) = ABC + A'BC + AB'C'*

*F(A, B, C)* in this case could be described as a standard sum-of-products (*SOP*) function according to the analogy that exists between *OR* and addition (+), and between *AND* and product (.); the *NOT* operation is indicated by an Apostrophe " ' " following the variable name. Usually, standard representations are also referred to as canonical representations.

In an alternative formulation, consider the following function *E(A,B,C)* in a product-of-sums (*POS*) form:

$E(A, B, C) = (A + B' + C).(A' + B + C)( A + B + C')$

The canonical *POS* implementation is shown in Figure 6.b. Some other specifications might require functions with a greater number of inputs and accordingly more complicated design process.

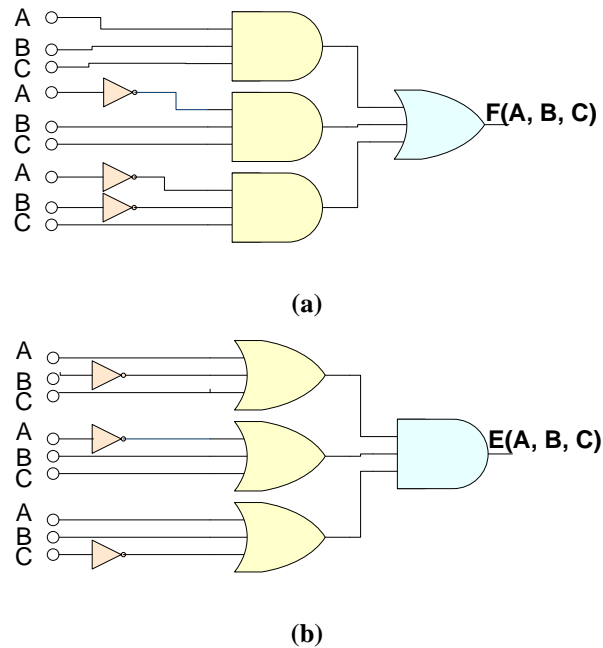

(a)

(b)

**Figure 6. AND-OR-Inverter implementation of the function (a) SOP: F(A, B, C) = ABC + A'BC + AB'C'. (b) POS: E(A, B, C) = (A + B' + C).(A' + B + C)( A + B + C')**

The complexity of a digital logic circuit that corresponds to a *Boolean* function is directly related to the complexity of the base algebraic function. *Boolean* functions may be simplified by several means. The simplification process that produces an expression with the least number of terms with the least number of variables is usually called minimization. The minimization has direct effects on reducing the cost of the implemented circuit and sometimes enhancing its performance. The minimization (optimization) techniques range from simple (manual) to complex (automated). An example of manual optimization methods is the *Karnough* map (*K*-map).

## K-maps

A *K*-map is like a truth table as it presents all the possible values of input variables and their corresponding output. The main difference between *K*-maps and truth tables is in the cell's arrangement. In a *K*-map, Cells are arranged in a way so that simplification of a given algebraic expression is simply a matter of properly grouping the cells. *K*-maps can be used for expressions with different number of input variables; three, four, or five. In the following examples, maps with only three and four variables are shown to stress the principle. Methods for optimizing expressions with more than five variables can be found in the literature. The *Quine-McClusky* is an example method that can accommodate several variables larger than five (2).

A 3-variable *K*-map is an array of *8* (or $2^3$) cells. Figure 7.a depicts the correspondence between three inputs (*A*, *B*, and *C*) truth table and a *K*-map. The value of a given cell represents the output at certain binary values of *A*, *B*, and *C*. in a similar way, a 4-variable *K*-map is arranged as shown in Figure 7.b. *K*-maps could be used for expressions in either *POS* or *SOP* forms. Cells in a *K*-map are arranged so that they satisfy the *Adjacency* property; where only a single variable changes its value between adjacent cells. For instance, the cell *000*, that is the binary value of the term *A'B'C'*, is adjacent to cell *001* that corresponds to the term *A'B'C*. The cell *0011* (*A'B'CD*) is adjacent to the cell *0010* (*A'B'C'D*).

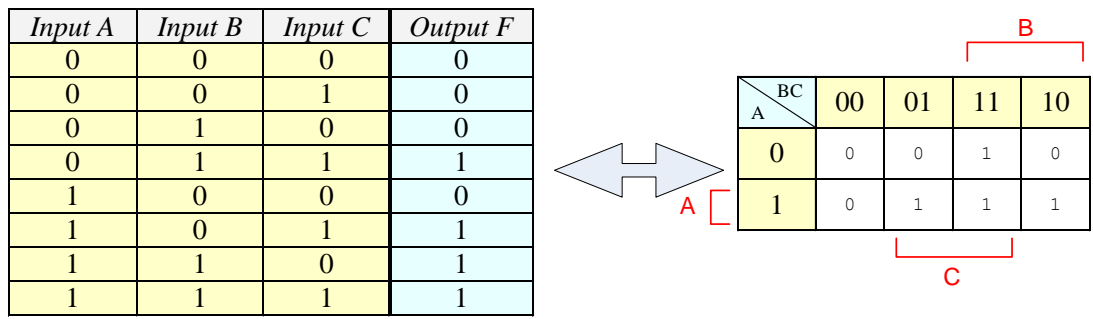

(a)

(b)

**Figure 7. (a) The correspondence between three inputs (*A*, *B*, and *C*) truth table and a *K-map*. (b) An empty 4-variable *K-map*.**

## Minimizing SOP Expressions

The minimization of an algebraic *Boolean* function *f* has the following four key steps:
1. Evaluation.
2. Placement.
3. Grouping.
4. Derivation.

The minimization starts by evaluating each term in the function *f* and then placing a *1* in the corresponding cell on the *K*-map. A term *ABC* in a function *f(A, B, C)* is evaluated to *111*, another term *AB'CD* in a function *g(A, B, C, D)* is evaluated to *1011*. An example of evaluating and placing the following function *f* is shown in Figure 8.a:

*f(A, B, C) = A'B'C' + A'B'C + ABC' + AB'C'*

After placing the *1s* on a *K*-map, grouping filled-with-*1s* cells is done according to the following rules (See Figure 8.b):

- A group of adjacent filled-with-*1s* cells must contain a number of cells that belongs to the set of powers of two (1, 2, 4, 8, or 16).
- A group should include the largest possible number of filled-with-*1s* cells.
- Each *1* on the *K*-map must be included in at least one group.
- *Cells* contained in a group could be shared within another group as long as overlapping groups included non-common *1s*.

After the grouping step, the derivation of minimized terms is done according to the following rules:
- Each group containing *1s* creates one product term.
- The created product term includes all variables that appear in only one form (completed or uncomplemented) across all cells in a group.

After deriving terms, the minimized function is composed of their sum. An example derivation is shown in Figure 8.b. Figure 9 presents the minimization of the following function:
*g(A, B, C, D) = AB'C'D' + A'B'C'D' + A'B'C'D' + A'B'CD + AB'CD + A'B'CD' +*
    *A'BCD + ABCD + AB'CD'*

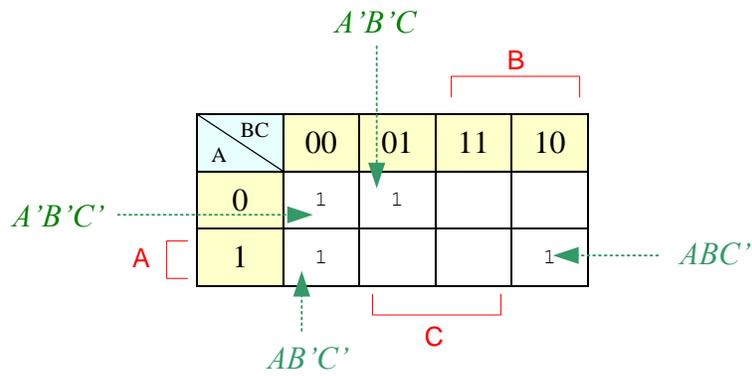

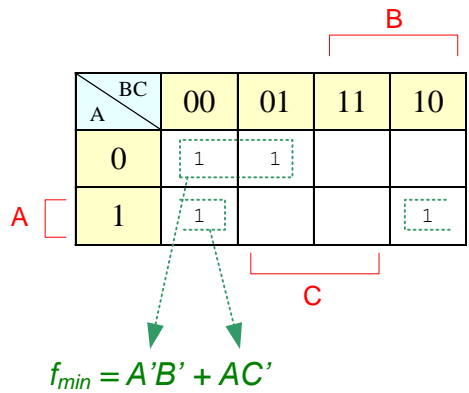

**Figure 8.** (a) Terms evaluation of the function $f(A, B, C) = A'B'C' + A'B'C + ABC' + AB'C'$. (b) Grouping and derivation.

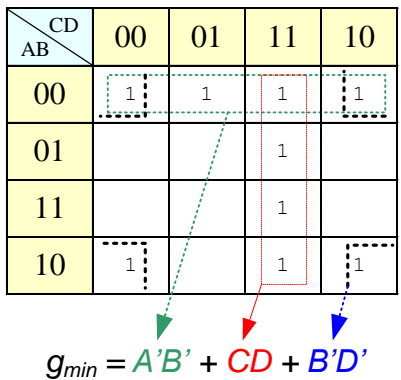

$g_{min} = A'B' + CD + B'D'$

**Figure 9.** minimization steps of the following function: $g(A, B, C, D) = AB'C'D' + A'B'C'D' + A'B'C'D' + A'B'CD + AB'CD + A'B'CD' + A'BCD + ABCD + AB'CD'$

## Combinational Logic Design

The basic combinational logic design steps could be summarized as follows:
1. Specification of the required circuit.
2. Formulation of the specification to derive algebraic equations.
3. Optimization (minimization) of the obtained equations
4. Implementation of the optimized equations using a suitable hardware (*IC*) technology.

The above steps are usually joined with an essential verification procedure which ensures the correctness and completeness of each design step.

As an example, consider the design and implementation of a 3 variables majority function. The function *F(A, B, C)* will return a *1* (*High* or *True*) whenever the number of *1s* in the inputs is greater than or equal to the number of *0s*.

The above specification could be reduced into a truth table as shown in Figure 7.a. The terms that make the function *F* return a *1* are the terms *F(0, 1, 1)*, *F(1, 0, 1)*, *F(1, 1, 0)*, or *F(1, 1, 1)*. This could be alternatively formulated as in the following equation:

$$F = A'BC + AB'C + ABC' + ABC$$

Following the specification and the formulation, a *K*-map is used to obtain the minimized version of *F* (called $F_{min}$). Figure 10.a depicts the minimization process. Figure 10.b shows the implementation of $F_{min}$ using standard *AND-OR-NOT* gates.

## Combinational Logic Circuits

Famous combinational circuits that are widely adopted in digital systems include Encoders, Decoders, Multiplexers, Adders, some Programmable Logic Devices (*PLDs*), etc. The basic operation of Multiplexers, Half-adders, and simple *PLDs* (*SPLDs*) is described in the following lines.

A multiplexer (*MUX*) selects one of *n* input lines and provides it on a single output. The select lines, denoted *S*, identify or address one of the several inputs. Figure 11.a shows the block diagram of a 2-to-1 multiplexer. The two inputs can be selected by one select line, *S*. If the selector *S = 0*, input line $d_0$ will be the output *O*, otherwise, $d_1$ will be produced at the output. An *MUX* implementation of the majority function *F(A, B, C)* is shown in Figure 11.b.

A half-adder inputs two binary digits to be added and produces two binary digits representing the sum and carry. The equations, implementation and symbol of a half-adder are shown in Figure 12.

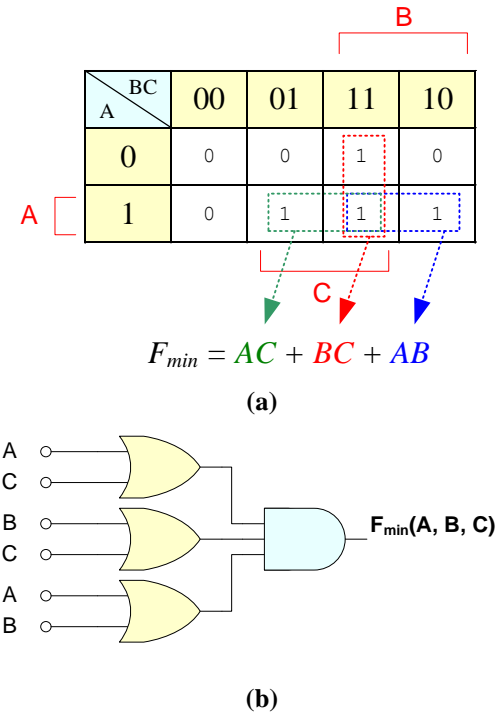

$$F_{min} = AC + BC + AB$$

(a)

(b)

**Figure 10. (a) Minimization of a 3 variables majority function. (b) Implementation of a minimized three variables majority function.**

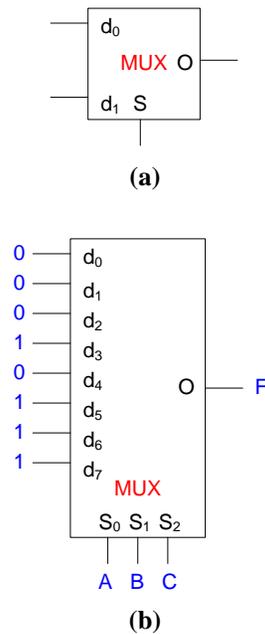

(a)

(b)

**Figure 11. (a) Minimization of a 3 variables majority function. (b) Implementation of a minimized three variables maj`ority function.**

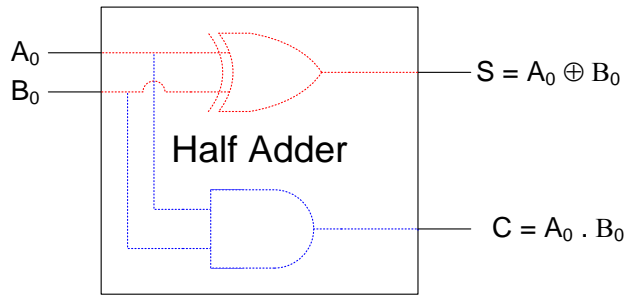

**Figure 12. The equations, implementation and symbol of a half-adder. The used symbol for a *XOR* operation is '⊕'.**

Simple *PLDs* (*SPLDs*) are usually built from combinational logic blocks with pre-routed wiring. In implementing a function on a *PLD*, the designer will only decide of which wires and blocks to use; this step is usually referred to as programming the device. Programmable Logic Array (*PLA*) and the Programmable Array Logic (*PAL*) are commonly used *SPLD*. A *PLA* has a set of programmable *AND* gates, which link to a set of programmable *OR* gates to produce an output (See Figure 13.a). A *PAL* has a set of programmable *AND* gates, which link to a set of fixed *OR* gates to produce an output (See Figure 13.b). The *AND-OR* layout of a *PLA/PAL* allows for implementing logic functions that are in an *SOP* form. A *PLA* implementation of the majority function *f(A, B, C)* is shown in Figure 13.c.

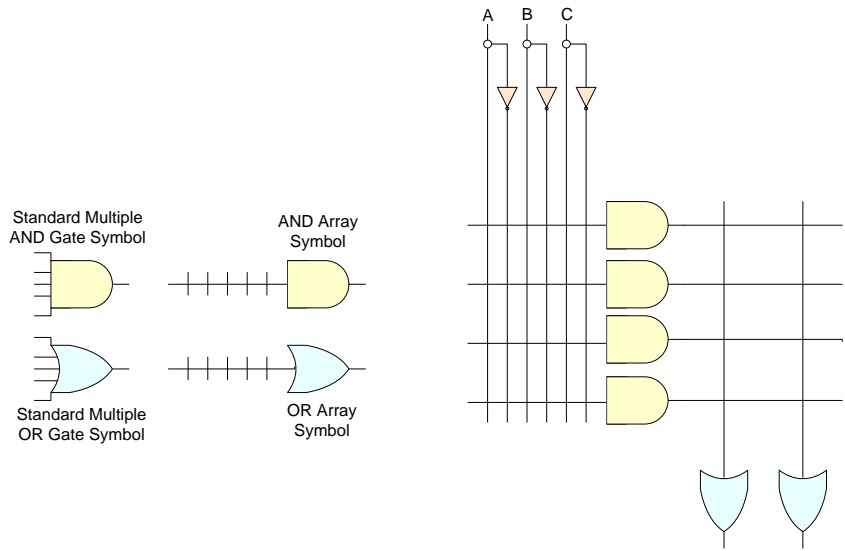

(a)

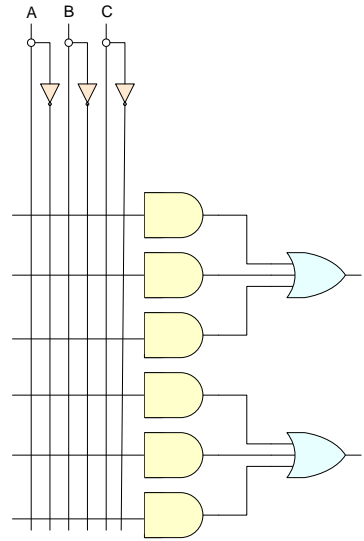

(b)

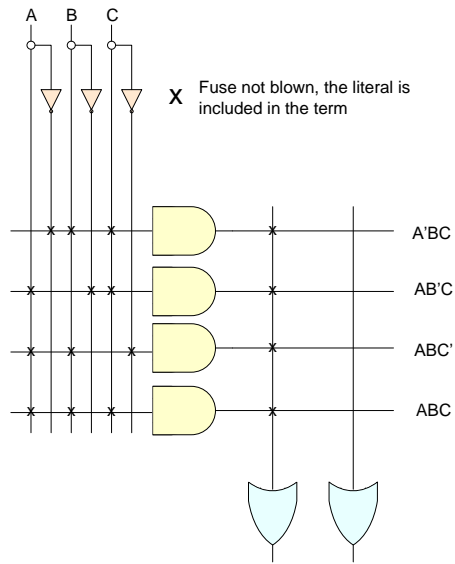

(c)

**Figure 13.** (a) A 3-input 2-output *PLA* with its *AND* Arrays and *OR* Arrays. An *AND* array is equivalent to a standard multiple-input *AND* gate, and an *OR* array is equivalent to a standard multiple-input *OR* gate. (b) A 3-input 2-output *PAL*. (c) A *PLA* implementation of the majority function *F(A, B, C)*

## Sequential Logic

In practice, most digital systems contain combinational circuits along with memory; these systems are known as sequential circuits. In sequential circuits, the present outputs depend on the present inputs and the previous states stored in the memory elements. Sequential circuits are of two types synchronous and asynchronous. In a synchronous sequential circuit, a clock signal is used at discrete instants of time to synchronize desired operations. A clock is a device that generates a train of pulses as shown in Figure 14. Asynchronous sequential circuits don't require synchronizing clock pulses; however, the completion of an operation signals the start of the next operation in sequence.

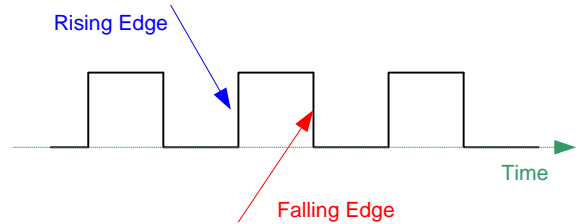

**Figure 14. Clock pulses.**

In synchronous sequential circuits, the memory elements are called flip-flops and can store only one bit. Arrays of flip-flops are usually used to accommodate for bit-width requirements of binary data. A typical synchronous sequential circuit contains a combinational part, sequential elements, and feedback signals coming from the output of the sequential elements.

## Flip-flops

Flip-flops are volatile elements, where the stored bit is stored if power is available. Flip-flops are designed using basic storage circuits called latches. The most common latch is the *SR* (Set to 1 - Reset to 0) latch. An *SR* latch could be formed with two cross-coupled *NAND* gates as shown in Figure 15. The responses to various inputs to the *SR* latch are setting *Q* to *1* for an *SR* input of *01* (*S* is active low, i.e. *S* is active when it is equal to *0*), resetting *Q* to *0* for an *SR* input of *10* (*R* here is also active low), memorizing the current state for an *SR* input of *11*. The *SR* input of *00* is considered invalid.

A flip-flop is a latch with a clock input. A flip-flop that changes state either at the positive (rising) edge or the negative (falling) edge of the clock is called an edge-triggered flip-flop (See Figure 14). The three famous edge-triggered flip-flops are the *RS*, *JK*, and *D* flip-flops.

An *RS* flip-flop is a clocked *SR* latch with two more *NAND* gates (See Figure 15.b). The symbol and the basic operation of an *RS* flip-flop are illustrated in Figure 16.a. The operation of an *RS* flip-flop is different from that of an *SR* latch and responds differently to different values of *S* and *R*. The *JK* and *D* flip-flops are derived from the *SR* flip-flop. However, the *JK* and *D* flip-flops are more widely used (2). The *JK* flip-flop is identical to the *SR* flip-flop with a single difference, where it has no invalid state (See Figure 16.b). The *D* flip-flop has only one input formed with an *SR* flip-flop and an inverter (See Figure

16.c); thus, it only could set or reset. The *D* flip-flop is also known as transparent flip-flop, where output will have the same value of the input after one clock cycle.

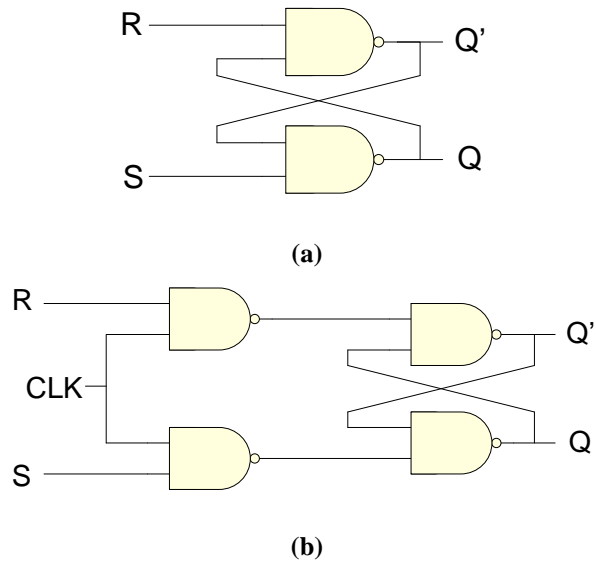

(a)

(b)

**Figure 15. (a) An SR latch. (b) an *RS* flip-flop.**

| S | R | Next State Q(t + 1) | |
|---|---|---|---|
| 0 | 0 | Q(t) | Unchanged |
| 0 | 1 | 0 | Reset |
| 1 | 0 | 1 | Set |
| 1 | 1 | - | Invalid |

(a)

| J | K | Next State Q(t + 1) | |
|---|---|---|---|
| 0 | 0 | Q(t) | Unchanged |
| 0 | 1 | 0 | Reset |
| 1 | 0 | 1 | Set |
| 1 | 1 | Q'(t) | Complement |

(b)

| D | Next State Q(t + 1) | |
|---|---|---|
| 0 | 0 | Reset |
| 1 | 1 | Set |

(c)

**Figure 16. (a) The symbol and the basic operation of (a) *RS* flip-flop (b) *JK* flip-flop (c) *D* flip-flop.**

# Sequential Logic Design

The basic sequential logic design steps are generally identical to those for combinational circuits; these are Specification, Formulation, Optimization, and the Implementation of the optimized equations using a suitable hardware (*IC*) technology. The differences between sequential and combinational design steps appear in the details of each step.

The specification step in sequential logic design usually describes the different states the sequential circuit goes through. A typical example for a sequential circuit is a counter that undergoes 8 different states, for instance, 0, 1, 2, 3, up till 7. A classic way to describe the state transitions of sequential circuits is a state diagram. In a state diagram, a circle represents a state, and an arrow represents a transition. The proposed example assumes no inputs to control the transitions among states. Figure 17.a shows the state diagram of the specified counter. The number of states determines the minimum number of flip-flops to be used in the circuit. In the case of the *8*-states counter, the number of flip-flops should be *3*; in accordance with the formula *8* equals $2^3$. At this stage, the states could be coded in binary. For instance, the stage representing count *0* is coded to binary *000*; the stage of count *1* is coded to binary *001*, etc.

The state diagram is next to be described in a truth table style, usually known as state table, from which the formulation step could be carried forward. For each flip-flop an input equation is derived (See Figure 17.b). The equations are then minimized using *K-maps* (See Figure 17.c). The minimized input equations are then implemented using a suitable hardware (*IC*) technology. The minimized equations are then to be implemented (See Figure 17.d).

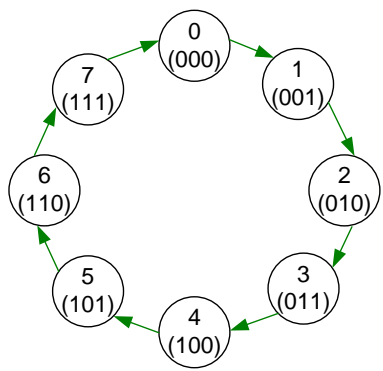

(a)

| Present State | | | Next State | | |
|---|---|---|---|---|---|
| $Q_{A(t)}$ | $Q_{B(t)}$ | $Q_{C(t)}$ | $Q_{A(t+1)}$ | $Q_{B(t+1)}$ | $Q_{C(t+1)}$ |
| 0 | 0 | 0 | 0 | 0 | 1 |
| 0 | 0 | 1 | 0 | 1 | 0 |
| 0 | 1 | 0 | 0 | 1 | 1 |
| 0 | 1 | 1 | 1 | 0 | 0 |
| 1 | 0 | 0 | 1 | 0 | 1 |
| 1 | 0 | 1 | 1 | 1 | 0 |
| 1 | 1 | 0 | 1 | 1 | 1 |
| 1 | 1 | 1 | 0 | 0 | 0 |

(b)

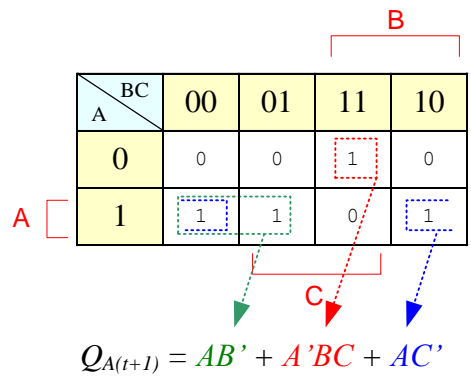

$Q_{A(t+1)} = AB' + A'BC + AC'$

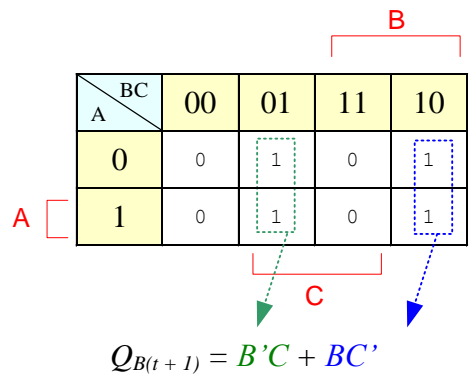

$Q_{B(t+1)} = B'C + BC'$

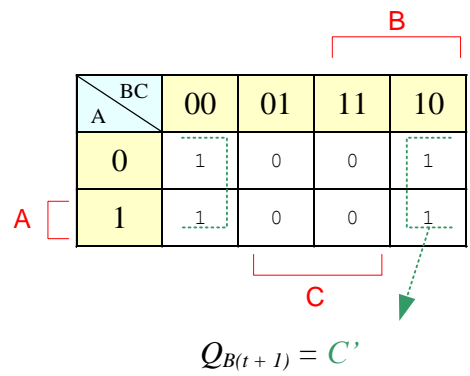

$Q_{B(t+1)} = C'$

(c)

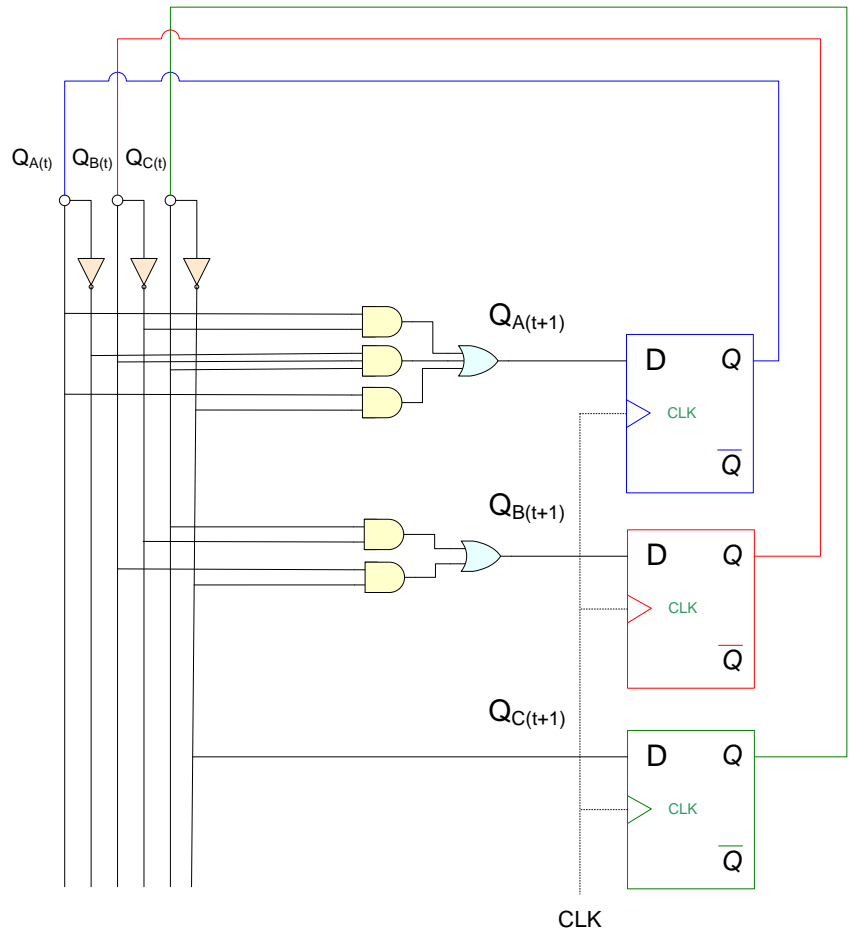

**(d)**

**Figure 17. (a) The state diagram of the specified counter. (b) The state table. (c) Minimization of input equations. (d) Implementation of the counter.**

## Modern Logic Design

The task of manually designing hardware tends to be extremely tedious, and sometimes impossible, with the increasing complexity of modern digital circuits. Fortunately, the demand on large digital systems has been accompanied with a fast advancement in *IC* technologies. Indeed, *IC* technology has been growing faster than the ability of designers to produce hardware designs. Hence, there has been a growing interest in developing techniques and tools that facilitate the process of logic design.

Two different approaches emerged from the debate over ways to automate hardware logic design. On one hand, the capture-and-simulate proponents believe that human designers have good design experience that cannot be automated. They also believe that a designer can build a design in a bottom-up style from elementary components such as transistors and gates. Since the designer is concerned with the deepest details of the design, optimized and cheap designs could be produced. On the other hand, the describe-and-synthesize advocates believe that synthesizing algorithms can out-perform human designers. They also believe that a top-down fashion would be better suited for designing complex systems. In describe-and-synthesize methodology, the designers firstly describe the design. Then, computer aided design (*CAD*) tools can generate the physical and electrical structure. This approach describes the intended designs using special languages called hardware description languages (*HDLs*). Some *HDLs* are very similar to traditional programming languages like *C*, *Pascal*, etc. (3). *Verilog* (4) and *VHDL* (Very High Speed Integrated Circuit Hardware Description Language) (5) are by far the most commonly used *HDLs* in industry.

Hardware synthesis is a general term used to refer to the processes involved in automatically generating a hardware design from its specification. High-level Synthesis (*HLS*) could be defined as the translation from a behavioral description of the intended hardware circuit into a structural description. The behavioral description represents an algorithm, equation, etc., while a structural description represents the hardware components that implement the behavioral description.

The chained synthesis tasks at each level of the design process include system synthesis, register-transfer synthesis, logic synthesis, and circuit synthesis. System synthesis starts with a set of processes communicating though either shared variables or message passing. Each component can be described using a register-transfer language (*RTL*). *RTL* descriptions model a hardware design as circuit blocks and interconnecting wires. Each of these circuit blocks could be described using *Boolean* expressions. Logic synthesis translates *Boolean* expressions into a list of logic gates and their interconnections (netlist). Based on the produced netlist, circuit synthesis generates a transistor schematic from a set of input-output current, voltage and frequency characteristics or equations.

The Logic synthesis step automatically converts a logic-level behavior, consisting of logic equations and/or a Finite State Machines (*FSMs*), into a structural implementation (3). Finding an optimal solution for complex logic minimization problems is very hard. As a consequence, most logic synthesis tools use heuristics. A heuristic is a technique whose result can hopefully come close to the optimal solution. The impact of complexity and of the use of heuristics on logic synthesis is significant. Logic synthesis tools differ tremendously according to the heuristics they use. Some computationally-intensive heuristics requires long run times and thus powerful workstations producing high-quality solutions. However, other logic synthesis tools use fast heuristics that are typically found on personal computers (*PCs*) producing solutions with less quality. Tools with expensive heuristics usually allow a user to control the level of optimization to be applied.

Continuous efforts have been made paving the way for modern logic design. These efforts included the development of many new techniques and tools. An approach to logic minimization using a new sum operation called multiple valued *EXOR* is proposed in (6) based on neural computing.

In (7), Tomita et al. discuss the problem of locating logic design errors, and proposes an algorithm to solve it. Based on the results of logic verification, the authors introduce an input pattern for locating design errors. An algorithm for locating single design errors with the input patterns has been developed.

Efforts for creating tools with higher levels of abstraction in design lead to the production of many powerful modern hardware design tools. Ian Page and Wayne Luk developed a compiler that transformed a subset of *Occam* into a netlist (8). Nearly ten years later we have seen the development of *Handel-C*, the first commercially available high-level language for targeting programmable logic devices (such as field programmable gate arrays - *FPGAs*) (9).

A prototype *HDL* called Lava is developed by Satnam Singh at *Xilinx* and Mary Sheeran and Koen Claessen at Chalmers University in Sweden (10). Lava allows circuit tiles to be composed using powerful high-order combinators. This language is embedded in the *Haskell* lazy functional programming language. *Xilinx* implementation of *Lava* is designed to support the rapid representation, implementation and analysis of high performance *FPGA* circuits.

Logic design has an essential impact on the development of modern digital systems. In addition, logic design techniques are a primary key in various modern areas, such as, embedded systems design, reconfigurable systems (11), hardware/software co-design, etc.

## Bibliography.

[1] F. Vahid et al., Embedded System Design: A Unified Hardware/Software Introduction, New York: John Wiley & Sons, 2002.
[2] T. Floyd , Digital Fundamentals with PLD Programming, New Jersey: Prentice Hall, 2006.
[3] S. Hachtel, Logic Synthesis and Verification Algorithms, Norwell: Kluwer, 1996.
[4] IEEE, *Verilog HDL language reference manual*, IEEE Standard 1364, 1995.
[5] IEEE, *Standard VHDL reference manual*, IEEE Standard 1076, 1993.
[6] A. Hozumi, N. Kamiura, Y. Hata, K. Yamato, Multiple-valued logic design using multiple-valued EXOR, *Proc. Multiple-Valued Logic*, 290 - 294, 1995.

[7] M. Tomita, H. Jiang, T. Yamamoto, Y. Hayashi, An algorithm for locating logic design errors, *Proc. Computer-Aided Design*, 468 – 471, 1990.
[8] I. Page and W. Luk, Compiling Occam into field-programmable gate arrays, *Proc. Workshop on Field Programmable Logic and Applications*: 271–283, 1991.
[9] S. Brown and J. Rose, Architecture of FPGAs and CPLDs: A Tutorial, IEEE Design and Test of Computers, 2: 42–57, 1996.
[10] K. Claessen. *Embedded Languages for Describing and Verifying Hardware*. PhD Thesis, Chalmers University of Technology and Göteborg University, 2001.
[11] E. Mirsky and A. DeHon, MATRIX: A reconfigurable computing architecture with configurable instruction distribution and deployable resources, *Proc. IEEE Workshop on FPGAs for Custom Computing Machines*, 157 – 166, 1996.

## Reading List

T. Floyd , Digital Fundamentals with PLD Programming, New Jersey: Prentice Hall, 2006.

M. Mano et al., Logic and Computer Design Fundamentals, New Jersey: Prentice Hall, 2004.

## Cross-references

Programmable Logic Devices, See *PLDs*.
Synthesis, See High-level Synthesis
Synthesis, See Logic Synthesis